\documentclass[
    ,final            
  ]
  {aipproc}

\layoutstyle{6x9}

\newcommand{\Msol}{\rm\,M_{\odot}}

\newcommand{\rh}{\rm\,r_{h}}

\newcommand{\Mpc}{\rm\,Mpc}

\newcommand{\kms}{\rm\,km\,s^{-1}}
\newcommand{\Mr}{\rm\,M_{r}}
\newcommand{\fr}{\rm\,f_{red}}
\newcommand{\dfr}{\rm\,\Delta(f_{red})}
\newcommand{\mur}{\rm\,\mu_{red}}

\newcommand{\mub}{\rm\,\mu_{blue}}

\newcommand{\sigmar}{\rm\,\sigma_{red}}

\newcommand{\sigmab}{\rm\,\sigma_{blue}}

\newcommand{\driro}{\rm\,\Sigma_{r_i,r_o}}
\newcommand{\dsca}{\rm\,\Sigma_{0,0.5}}

\newcommand{\dscc}{\rm\,\Sigma_{0.5,1}}

\newcommand{\dsce}{\rm\,\Sigma_{2,3}}

\newcommand{\rdiv}{\rm\,r_{div}}

\begin{document}

\title{A multiscale approach to environment}

                

\author{David Wilman}{
  address={Max-Planck-Institut f\"ur Extraterrestrische Physik, Giessenbachstra\ss e, D-85748 Garching, Germany.}
}

\author{Stefano Zibetti}{
  address={Max-Planck-Institut f\"ur Astronomie, K\"onigstuhl 17, D-69117 Heidelberg, Germany.}
}

\author{Tam\'as Budav\'ari}{
  address={Department of Physics and Astronomy, The Johns Hopkins University, 3701 San Martin Drive, Baltimore, MD 21218, USA}
}

\begin{abstract}

Physical processes influencing the properties of galaxies can be traced by the dependence and evolution of galaxy properties on their environment. 
A detailed understanding of this dependence can only be gained through comparison of observations with models, with an appropriate quantification of the rich parameter space describing the environment of the galaxy. 
We present a new, multiscale parameterization of galaxy environment which retains an observationally motivated simplicity whilst utilizing the information present on different scales. 
We examine how the distribution of galaxy (u-r) colours in the Sloan Digital Sky Survey (SDSS), parameterized using a double gaussian (red plus blue peak) fit, depends upon multiscale density. 
This allows us to probe the detailed dependence of galaxy properties on environment in a way which is independent of the halo model.
Nonetheless, cross-correlation with the group catalogue constructed by Yang et al, 2007 shows that galaxy properties trace environment on different scales in a way which mimics that expected within the halo model.
This provides independent support for the existence of virialized haloes, and important additional clues to the role played by environment in the evolution of the galaxy population.
This work is described in full by Wilman~et~al, 2010, MNRAS, in press.

\end{abstract}

\maketitle


\section{Introduction}

The evolution of galaxies is heavily intertwined with the growth of the dark matter dominated structure in which they live, as baryons react to their local gravitational potential. 
This leads to a strong and evolving dependence of galaxy properties on their environment \citep[e.g.][]{Dressler80,Balogh04}. 
Environment is often defined in a observationally practical fashion, by counting the number of nearby neighbours. A different strategy is to assign galaxies to groups within a theoretical halo framework. 
Motivated by the need for a fair comparison of data with models, it would be preferable to define a statistical framework which retains the information contained by environment measured on different scales, without resorting to the inevitable assumptions required by a halo (group) finder during membership assignment. 
Simple two point galaxy statistics (such as correlation functions) trace the mean overdensity on different scales. However these are themselves closely correlated with one another. 
A better approach would be to examine how galaxy properties correlate with overdensity on one scale for a fixed overdensity on another, independently computed scale. 
Now that such a large, homogeneous sample of galaxies is available with the SDSS, this is now statistically possible \citep{Kauffmann04,Balogh04Ha,Blanton06,Blanton07}. 
In Wilman et al, 2010, we take the logical next step: defining a framework for quantifiable comparisons with models. The basic principles and initial results are described here.

\section{Data and Method}

\begin{figure}
  \centerline{\includegraphics[width=0.35\textwidth]{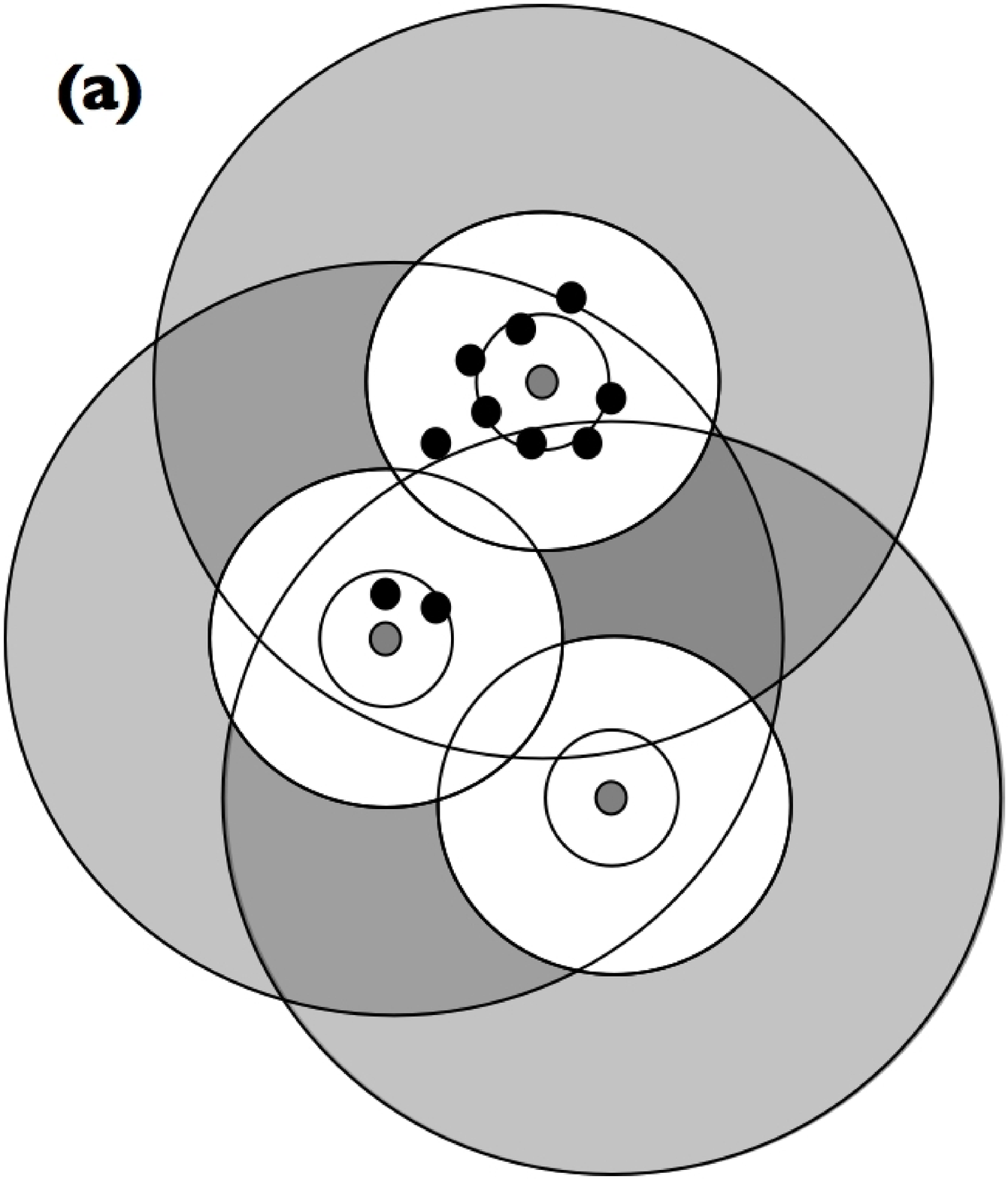}
              \hspace{0.02\textwidth}
              \includegraphics[width=0.55\textwidth]{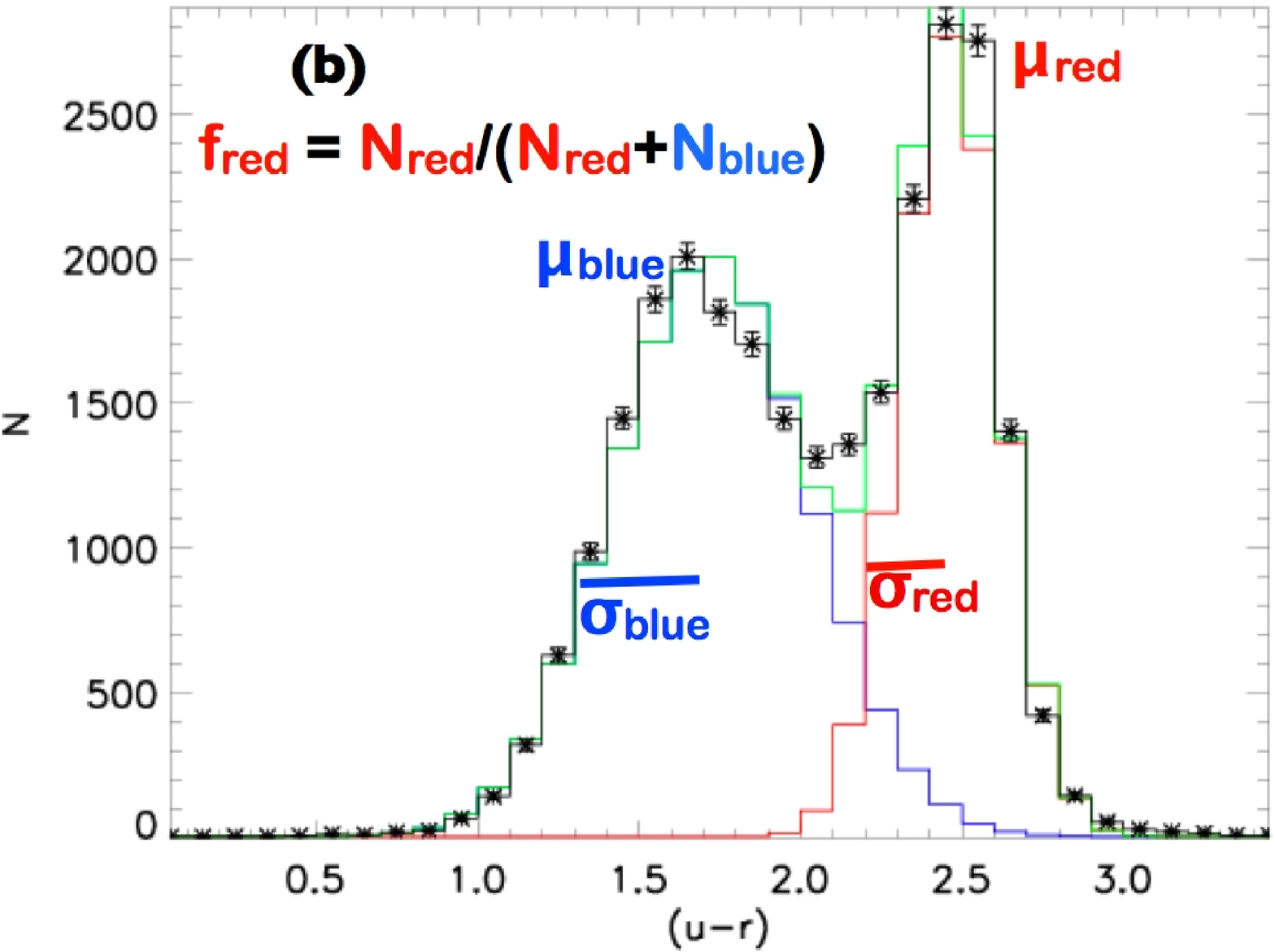}}
  \caption{(a) Illustrates the annuli within which density is computed on different scales, as applied to three galaxies in different environments. On the three scales measured (starting with the smallest), these galaxies have 0,0,5 neighbours (lowest galaxy), 2,0,8 neighbours (middle galaxy) and 3,5,3 neighbours (top galaxy). 
  (b) Demonstrating the bimodal distribution of SDSS galaxy (u-r) ``model'' colours. The colour distribution is well fit by a double gaussian distribution, described by five physically meaningful parameters ($\fr$, $\mur$, $\mub$, $\sigmar$, $\sigmab$).}
  \label{figure:annulibimodal}
\end{figure}

Our approach is simple and provides a method to study the density-dependence of galaxy properties on different scales. 
For each galaxy the density of $\Mr\leq-20.$ neighbours $\driro$ is computed within a cylindrical annulus of  depth $dv_{rest}\pm1000\kms$ and inner,outer radii $r_i$,$r_o$.  
This is computed on different scales, which, as illustrated in figure~\ref{figure:annulibimodal} (a), traces different aspects of environment. 

Our sample is volume-limited within SDSS DR5: $0.015\leq z\leq0.08$, $\Mr\leq-20.$, photometric completeness within $3\Mpc$ of the galaxy $\geq 98.645\%$, providing 73662 galaxies. 
For each galaxy, on each scale, a spectroscopic completeness factor is computed (the fraction of potential spectroscopic targets with redshifts) and used to correct the measurements of density. 

To see how galaxy properties depend upon multiscale density, we turn to the bimodal (u-r) colour distribution of SDSS galaxies. 
It has been shown that this is well described by a double gaussian fit, and that the resultant parameters are strong functions of galaxy luminosity or mass and local density \citep{Baldry04,Balogh04,Baldry06}. 
Figure~\ref{figure:annulibimodal} (b) shows an example of the double gaussian fit, illustrating the five independent parameters describing the distribution: the fraction of red galaxies $\fr$, the mean colour of each peak $\mur$, $\mub$ and the widths of those peaks $\sigmar$, $\sigmab$. 
Operating with the simplifying assumption that the red galaxy population is dominated by galaxies in which star formation has been truncated provides these parameters with relatively simple physical interpretations in terms of the overall star formation history for the galaxy population. 

\section{Results and Discussion}

\begin{figure}
  \includegraphics[height=.45\textheight]{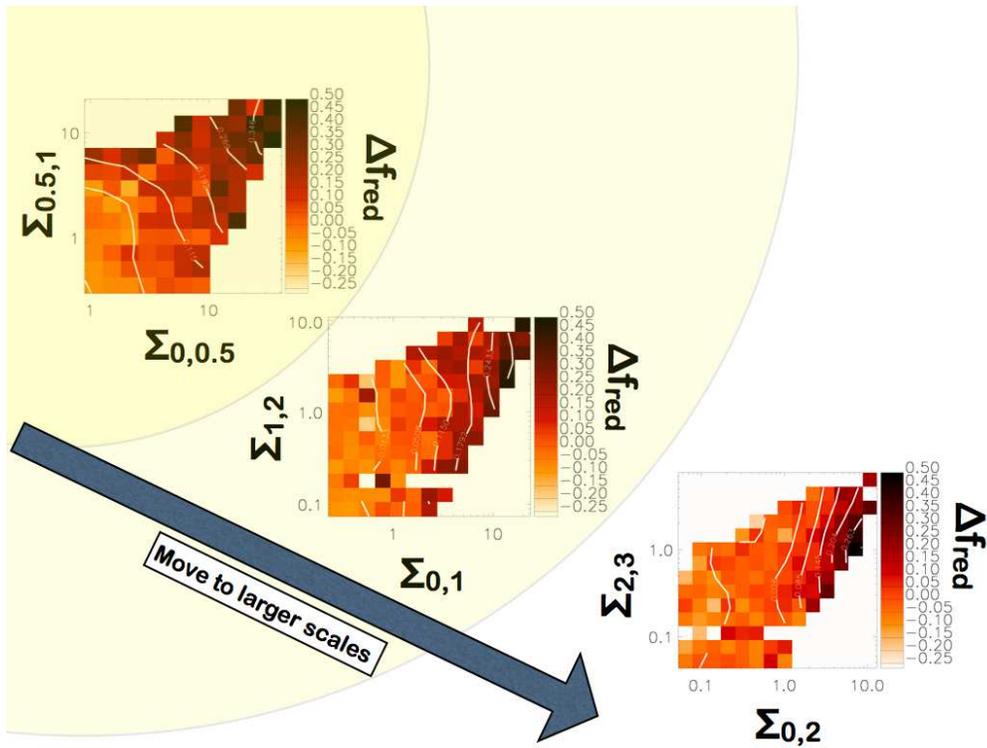}
  \caption{Dependence of the relative fraction of red galaxies, $\dfr$, on multiscale density. The bimodal fitting process is repeated independently for each bin in each two dimensional grid of interior and annulus density. This is illustrated for three choices of scales, moving from smaller to larger scales going from the top left to the bottom right. Grid colour changes and overplotted contours illustrate the direction of change within this parameter space. Interestingly our results illustrate a change from positive correlation with density (more red galaxies in denser environments) on small ($<1\Mpc$) scales, to an anti-correlation with density (less red galaxies in denser environments) on large ($>2\Mpc$) scales! This mimics the behaviour expected within a simple halo model framework (see text).}
  \label{figure:results}
\end{figure}

Our sample is divided into bins, spaced evenly in log-density within a grid of two-scale density. These scales are always chosen so that the smaller scale measures the density of neighbours up to a fixed distance (dividing radius $\rdiv$), and the larger scale then measures the density in some annulus beyond $\rdiv$. 
Figure~\ref{figure:results} shows the fraction of red galaxies resulting from bimodal fitting within each bin for $\rdiv=0.5$,$1.0$ and $2.0\Mpc$, where the outer radius for the outer annulus is $1.0$,$2.0$ and $3.0\Mpc$ respectively. 
There is also a (weak) dependence of $\fr$ on luminosity within this sample, and so we repeat this procedure for each $0.5$mag bin of luminosity in the range $-21.5\leq\Mr\leq-20.0$, with $\geq25$ galaxies per luminosity and density/density bin required to produce reasonable fits (reduced $\chi^2\sim1$). For each bin we compute a residual density dependent $\dfr=\fr(\Mr,\Sigma_{0,\rdiv},\Sigma_{\rdiv,r_o})-\fr(\Mr)$. 
The three luminosity bins (which have similar appearance) are then coadded ($\dfr$ is weighted by the number of contributing galaxies) to produce the final maps (figure~\ref{figure:results}). 
Colour gradients and contours illustrate the direction of change in $\dfr$ (vertical contours illustrate no large scale dependence). 

We find a changing dependence of $\dfr$ on density moving from small to large scales. 
On $<1\Mpc$ scales, there are more red galaxies in denser environments (a strong $\dsca$ dependence 
is supplemented with a residual positive correlation with $\dscc$). 
On larger scales there is no residual positive correlation, supporting previous results 
\citep{Kauffmann04,Blanton06,Blanton07}. 
However on still larger scales $>2\Mpc$, there is a highly significant {\it anti-correlation} 
with density at fixed $<2\Mpc$ scale density (i.e. more blue galaxies at higher $\dsce$)! 

The SDSS group (halo) catalogue of \citet{Yang07} is projected onto our multiscale density space. 
For the fraction of galaxies in haloes more massive than some fixed limit (e.g. $10^{12.5}\Msol$), 
we find a dependence on multiscale density remarkably similar to that seen in 
figure~\ref{figure:results}. 
i.e. positive correlations on small scales (at larger density, more galaxies in massive haloes) 
are replaced by anti-correlations on large scales (at larger density, fewer galaxies in massive haloes)!
More detailed analysis reveals the reason: 
For galaxies in the centre of haloes, density will always decrease with scale $r$, on average. 
Galaxies located a distance $\rh$ from the nearest halo centre will instead show increasing density 
with $r$ up to $r=\rh$ (where the annulus encompasses the halo core), and then decrease to larger $r$. 
i.e. The galaxies at highest density at scale $r$, at fixed smaller scale density, have $r_i< r\sim\rh <r_o$. 
The virial radius of a $10^{12.5}\Msol$ halo is $\sim0.2\Mpc$, 
so larger scales can trace the cores of haloes to which our galaxy does not belong. 

On larger scales this effect only becomes more important, leading to a lower fraction of galaxies in massive haloes at high densities on large scales - i.e. the anti-correlations. 
Therefore the higher fraction of blue galaxies at higher large scale density can be explained within the context of the halo model: These galaxies are in the accretion regions, and remain blue until after they have been accreted onto more massive haloes.

This work is explained in more detail by Wilman et al., 2010, MNRAS, in press. 
Our multiscale density parameterization of the environment will provide an observationally motivated parameter space to examine the environment and the associated dependence of galaxy properties, and to make quantitative comparisons with expectations coming from different physical models of galaxy evolution.





\bibliographystyle{aipproc}   


\IfFileExists{\jobname.bbl}{}
 {\typeout{}
  \typeout{******************************************}
  \typeout{** Please run "bibtex \jobname" to optain}
  \typeout{** the bibliography and then re-run LaTeX}
  \typeout{** twice to fix the references!}
  \typeout{******************************************}
  \typeout{}
 }

\end{document}